\documentclass[runningheads]{llncs}

\usepackage[labelformat=simple]{subcaption}

\usepackage{graphicx}
\usepackage{xcolor}
\usepackage{amsmath}
\usepackage{amsfonts}
\usepackage{cite}
\usepackage{bm}
\usepackage{todonotes}
\usepackage{url}
\usepackage[misc,geometry]{ifsym}
\begin{document}
\title{Going Off-Grid: Continuous Implicit Neural Representations for 3D Vascular Modeling}
\titlerunning{Continuous INRs for 3D Vascular Modeling}
% If the paper title is too long for the running head, you can set
% an abbreviated paper title here
%
%\author{***}
%\authorrunning{***}
%\institute{***}

 \author{Dieuwertje Alblas \inst{1}\textsuperscript{(\Letter)} \and
 Christoph Brune\inst{1} \and \\
 Kak Khee Yeung\inst{2, 3}
 \and
 Jelmer M. Wolterink\inst{1}}
% %
 \authorrunning{D. Alblas et al.}
% % First names are abbreviated in the running head.
% % If there are more than two authors, 'et al.' is used.
% %
 \institute{Department of Applied Mathematics, Technical Medical Centre, \\ University of Twente, Enschede, The Netherlands\\
 \email{d.alblas@utwente.nl} \and Department of Surgery, Amsterdam UMC location Vrije Universiteit Amsterdam, \\ Amsterdam, The Netherlands \and Amsterdam Cardiovascular Sciences, Microcirculation,\\ Amsterdam, The Netherlands}
\maketitle              % typeset the header of the contribution
\begin{abstract}
Personalised 3D vascular models are valuable for diagnosis, prognosis and treatment planning in patients with cardiovascular disease. Traditionally, such models have been constructed with explicit representations such as meshes and voxel masks, or implicit representations such as radial basis functions or atomic (cylindrical) shapes. Here, we propose to represent surfaces by the zero level set of their signed distance function (SDF) in a differentiable implicit neural representation (INR). This allows us to model complex vascular structures with a representation that is implicit, continuous, light-weight, and easy to integrate with deep learning algorithms. We here demonstrate the potential of this approach with three practical examples. First, we obtain an accurate and watertight surface for an abdominal aortic aneurysm (AAA) from CT images and show robust fitting from as few as 200 points on the surface. Second, we simultaneously fit nested vessel walls in a single INR without intersections. Third, we show how 3D models of individual arteries can be smoothly blended into a single watertight surface. Our results show that INRs are a flexible representation with potential for minimally interactive annotation and manipulation of complex vascular structures.

% Due to the recent advances of deep learning in medical image segmentation, voxel masks have become the primarily used representation of 3D volumes. Prior to the deep learning era, the landscape of alternative representations was more diverse, including radial basis functions, spherical harmonics and implicit functions to represent 3D shapes. In this work, we demonstrate the advantages of representing shapes implicitly within the scope of vascular analysis. We show the versatility of this representation, as well as how they can be reconstructed from a sparse point cloud representation.

\end{abstract}

\section{Introduction}

%%% ALTERNATIEVE INTRODUCTIE
Accurate and patient-specific models of vascular systems are valuable for diagnosis, prognosis and treatment planning in patients with cardiovascular disease. Personalised vascular models might be used for stent-graft sizing in patients with abdominal aortic aneurysms~\cite{sobocinski2013benefits} or for computational fluid dynamics (CFD)~\cite{tran2021patient}, or for doctor-patient communication and shared decision-making ~\cite{swart2019shared}. 
However, extracting these models from medical image data can be cumbersome. Commercial software and open-source software packages~\cite{antiga2008image,lan2018re,arthurs2021crimson} traditionally rely on the construction of cylindrical models~\cite{shani1984splines} in three steps. First, the lumen centerline is identified for each vessel. Then, local cross-sectional contours are determined and used to construct a watertight mesh model using (spline) interpolation. Finally, polygon mesh models of multiple vessels are blended to obtain a connected vascular tree. In this approach, tortuosity of the centerline can cause self-intersections of the orthogonal contours, resulting in surface folding of the final mesh model~\cite{gansca2002self}. Moreover, smoothly connecting triangular meshes around bifurcations is challenging.
%Hence, this approach heavily relies on expert-indicated center lumen lines (CLLs), tubular models~\cite{shani1984splines}, and error-prone blending strategies on triangular meshes.

Deep learning has made great progress towards automatic model building from images~\cite{chen2020deep, litjens2019state}. However, popular convolutional neural network-based methods return 3D voxel masks. Because voxel masks merely discretize an underlying continuous shape, their quality heavily depends on the resolution of the image data, and they are not guaranteed to be contiguous. Hence, voxel masks typically require additional processing steps before use in, e.g., CFD. 
% Few exceptions exist in which anatomical priors are explicitly embedded to extract (tubular) meshes, e.g. \cite{Alblas/et/al/2021, wickramasinghe2020voxel2mesh}. However, 
% Simlatly, mesh parametrisations suffer from surface folding, and blending of multiple polygon meshes into a single 3D model is challenging. 
There is a need for a vascular model shape representation that is continuous, modular, and can be easily integrated with existing deep learning methods. 

\begin{figure}[t!]
    \centering
    \begin{subfigure}[t]{0.33\textwidth}
        \includegraphics[width = \textwidth]{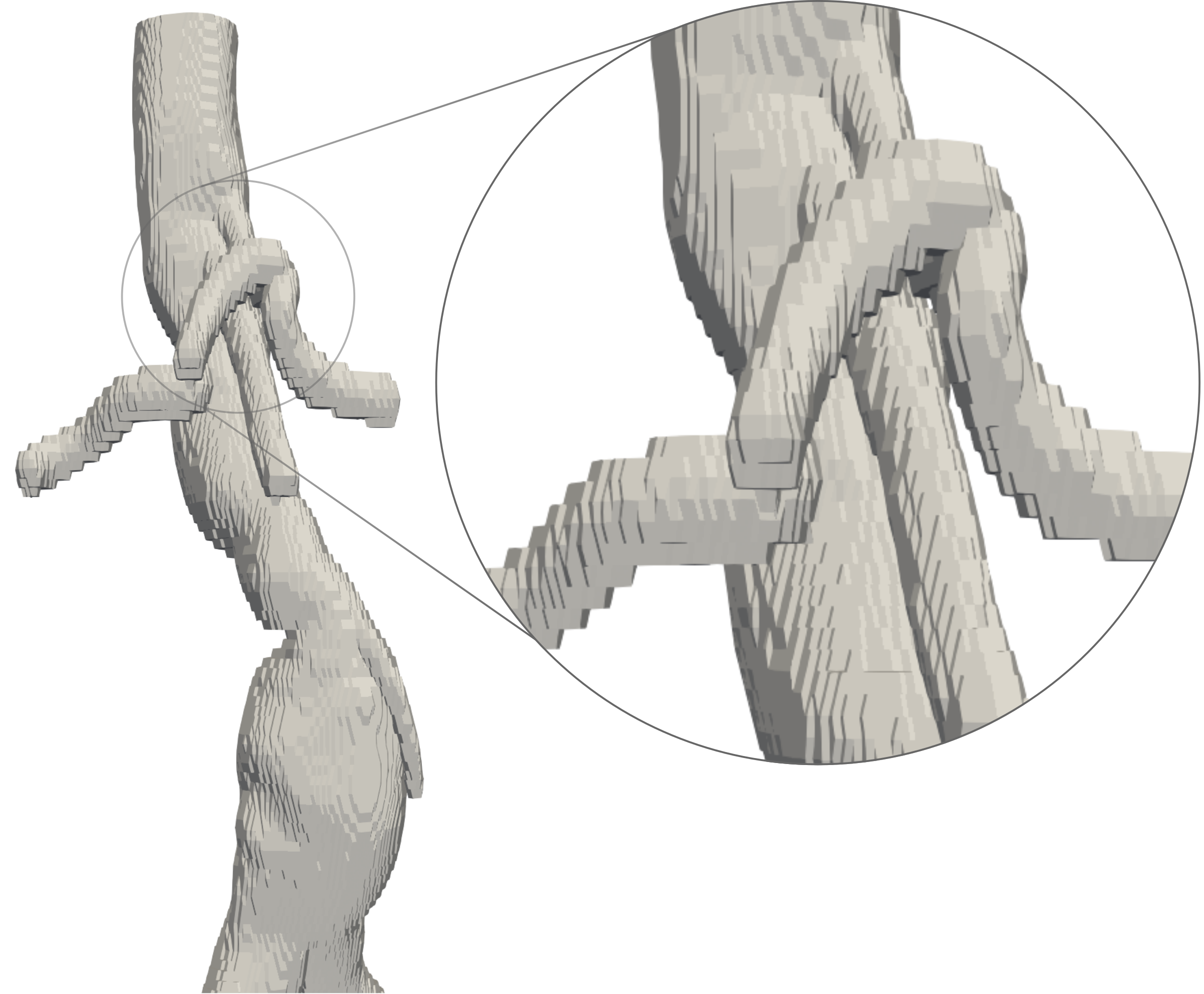}
        \caption{Voxel mask}
        \label{voxelmask}
    \end{subfigure}\begin{subfigure}[t]{0.33\textwidth}
    \includegraphics[width = \textwidth]{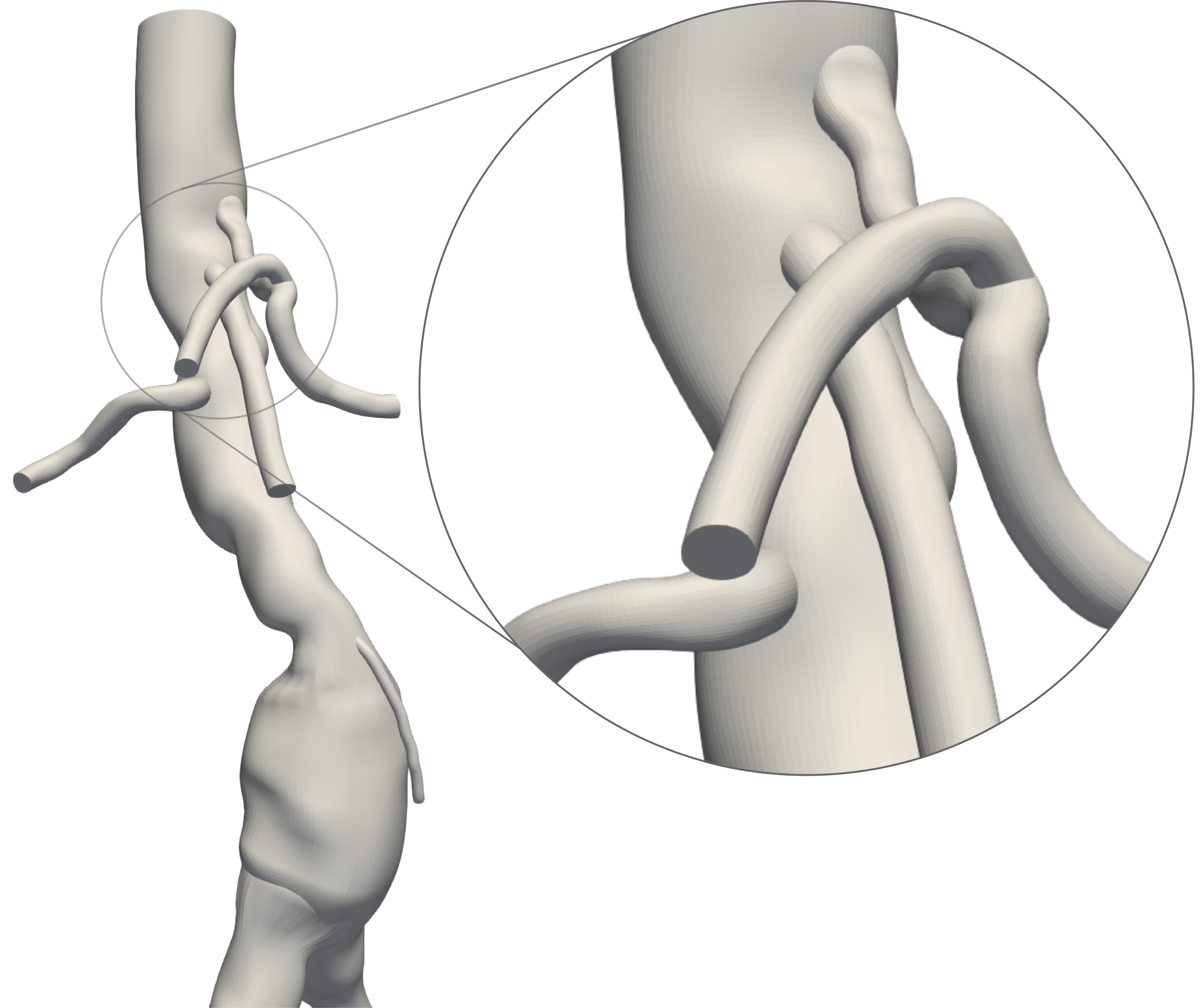}
    \caption{Mesh}
    \label{mesh}
    \end{subfigure}\begin{subfigure}[t]{0.33\textwidth}
    \includegraphics[width=\textwidth]{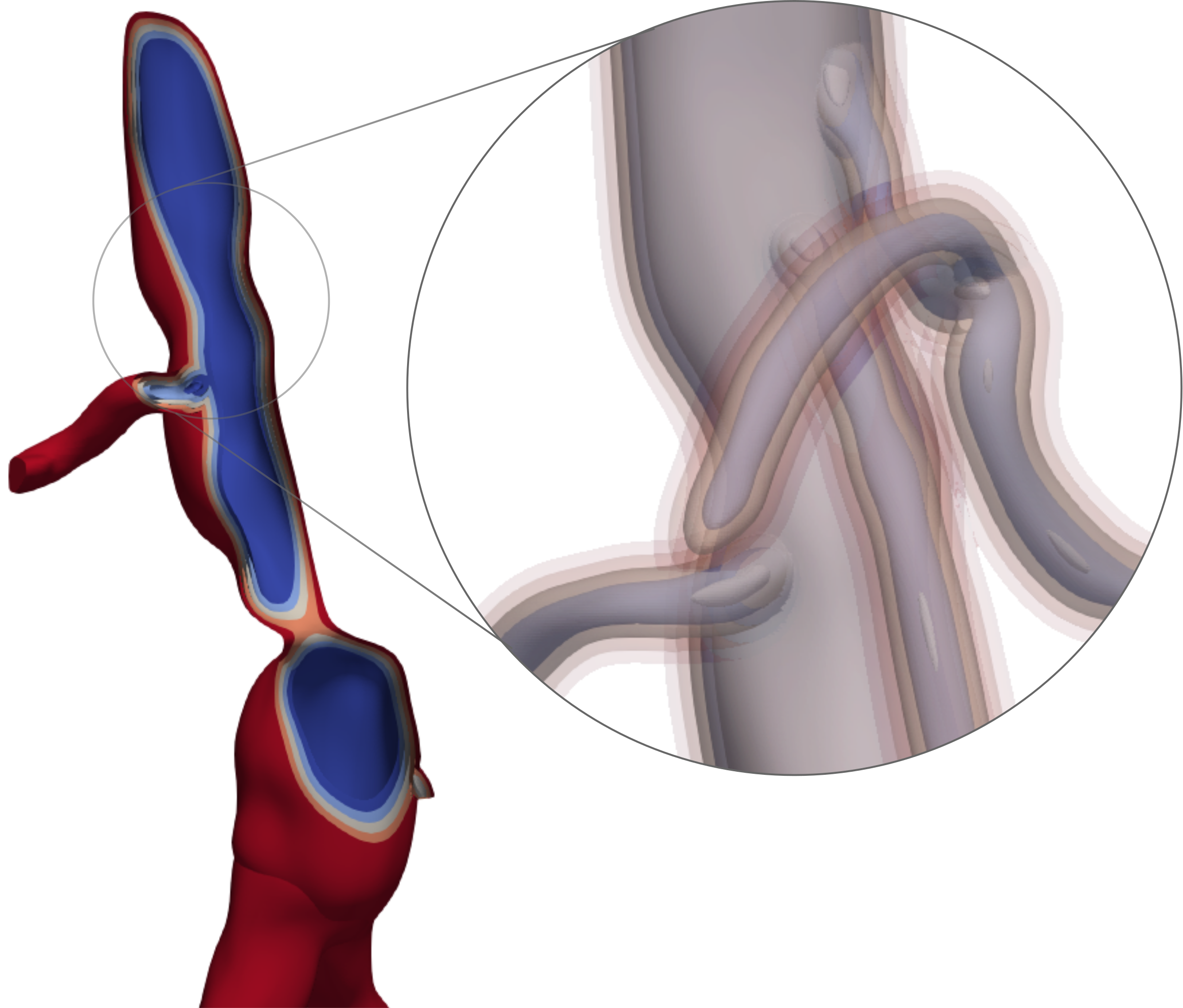}
    \caption{Signed distance function}
    \label{sdf}
    \end{subfigure}
    \caption{Different representations of a 3D aortofemoral tree \cite{wilson2013vascular}. \subref{voxelmask} and \subref{mesh} are \textit{explicit} representations; \subref{voxelmask} has non-smooth boundaries, whereas boundaries of \subref{mesh} are locally smooth. Both \subref{voxelmask} and \subref{mesh} are restricted to this resolution. \subref{sdf} \textit{implicitly} represents the surface with smooth boundaries, at any resolution.}
    \label{fig:representations}
\end{figure}
In this work, we adapt the work of Gropp et al. \cite{gropp2020implicit} to model vascular systems as combinations of level sets of signed distance functions represented in differentiable neural networks. 
% In particular, level sets of signed distance functions (SDFs) have several attractive properties that make them easy to (smoothly) combine and visualize at arbitrary resolution. 
Implicit representations and level sets have a substantial history in both segmentation~\cite{van2002level, lorigo2001curves} and 3D modeling \cite{hong2020high, kretschmer2013interactive} of vascular structures. Recently, there have been significant advances in signal representations using neural networks, i.e., implicit neural representations (INRs)\cite{mildenhall2020nerf, sitzmann2019siren, park2019deepsdf, martel2021acorn}. % Implicit neural representations can be directly optimized as part of any deep learning method. 
Continuous implicit neural representations of the signed distance function can be easily transformed into any explicit representation, while conversion between explicit representations is non-trivial (Fig. \ref{fig:representations}). In this paper, we show how a continuous representation can be obtained from an explicit one.
Moreover, an INR has a limited memory footprint that is independent of resolution, while the memory requirements of voxel masks and triangular meshes grow rapidly with an increase in resolution.

We here demonstrate that INRs are a potentially valuable tool to bridge the gap between vascular modeling on the one hand and deep learning on the other hand. First, we show how INRs can be used to reconstruct an implicit surface from a sparse point cloud and form an alternative to conventional annotation procedures. We evaluate the efficiency and robustness of this approach. 
% By omitting connected contours, we avoid self-intersections. 
Secondly, we show that we can use a single INR to represent multiple surfaces, and demonstrate the effectiveness on nested shapes in an AAA case study. 
Finally, we demonstrate the added value of implicit shape representations in the smooth blending of separate structures in the reconstruction of an aortofemoral tree. 

\section{Method}
We represent the surface of a vascular structure by the zero level set of its signed distance function (SDF), which we embed in a multilayer perceptron as an implicit neural representation.

\subsection{Signed distance functions}
A surface can be \textit{implicitly} described by the zero level set of its signed distance function (SDF).  In the case of a 2D surface $\mathcal{M}$ embedded in a 3D domain,  $SDF_{\mathcal{M}}(\bm{x}): \mathbb{R}^3 \to \mathbb{R}$ is defined as:
\begin{align}
    & SDF_\mathcal{M}(\bm{x}) = \begin{cases}
    -d(\bm{x}, \mathcal{M}) \hspace{0.5cm} & \bm{x} \text{ inside } \mathcal{M} \label{def_SDF}\\
    0 & \bm{x} \text{ on } \mathcal{M}\\
    d(\bm{x}, \mathcal{M}) & \bm{x} \text{ outside } \mathcal{M},
    \end{cases}\\
    &\text{therefore: }\mathcal{M} = \{ \bm{x} \in \mathbb{R}^3 | SDF_\mathcal{M}(\bm{x}) = 0 \}.
\end{align}
In Eq. \eqref{def_SDF}, $d(\bm{x},\mathcal{M}) = \min\limits_{\bm{y}\in \mathcal{M}} || \bm{x} - \bm{y} ||$, the minimal distance to the surface. An SDF satisfies the Eikonal equation, hence $||\nabla_x SDF_\mathcal{M}(\bm{x})|| = 1, \forall \bm{x}$. As this function is continuous on $\mathbb{R}^3$, the surface $\mathcal{M}$ is represented independently of resolution. 

%Equation \eqref{def_SDF} implies that $S$ is continuous and differentiable. As $SDF_S$ returns a distance, $|\nabla SDF_S| = 1$. One of the main benefits of using an SDF rather than an explicit representation is that it can represent an object in any arbitrary resolution, as $\phi$ is defined on the continuous domain and returns distances for all coordinates. The sign property simplifies determining whether a given point is inside or outside the shape. 

\begin{figure}[t!]
    \centering
    \begin{subfigure}[t]{0.49\textwidth}
    \includegraphics[width =\textwidth]{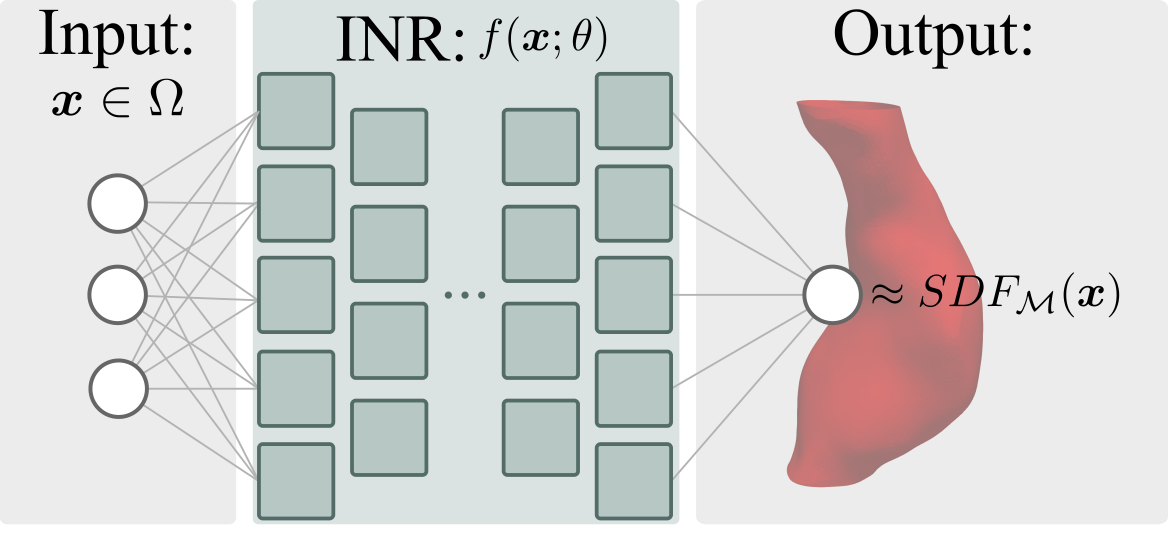}
    % \caption{INR parametrizing a single SDF.}
    \label{fig:INR_single}
    \end{subfigure}\hfill\begin{subfigure}[t]{0.49\textwidth}
    \includegraphics[width=\textwidth]{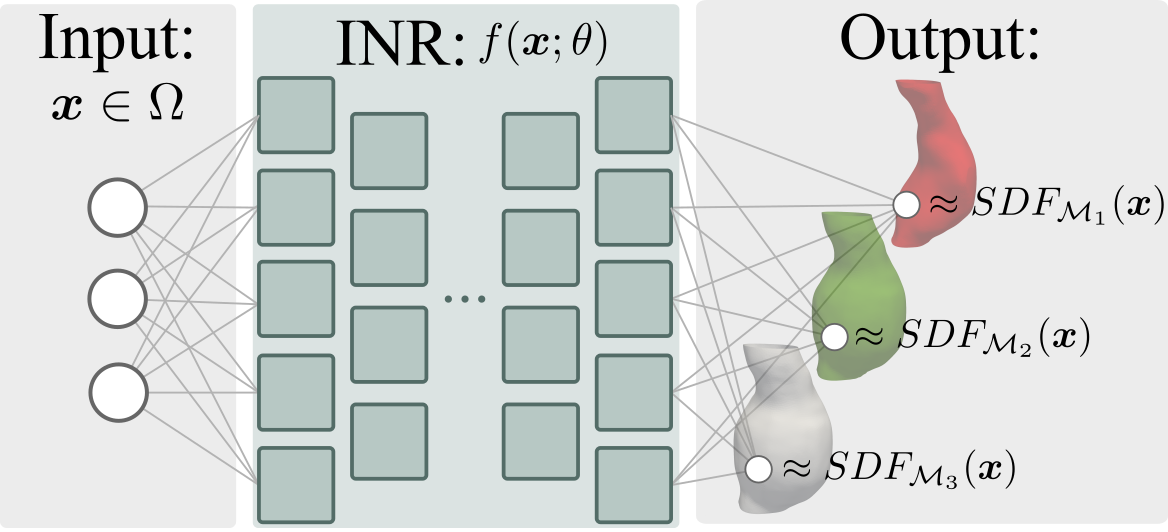}
    % \caption{INR parametrizing multiple SDF's.}
    \label{fig:INR_multiple}
    \end{subfigure}
    \caption{Schematic overview of an implicit neural representation, parametrising a single SDF \textit{(left)}, or three SDFs simultaneously \textit{(right)}. Coordinate values can be queried and the network returns local SDF values.}
    \label{fig:INR}
\end{figure}

\subsection{Implicit neural representations}
Signed distance functions are traditionally represented as a combination of SDFs of atomic shapes, or radial basis functions~\cite{carr1997surface}. However, a recent insight shows that functions in space can be approximated using neural networks \cite{park2019deepsdf, sitzmann2019siren}. These so-called implicit neural representations (INRs) have been widely used for, e.g., image representations \cite{chen2021learning}, novel view synthesis~\cite{mildenhall2020nerf}, image registration \cite{wolterink2021implicit}, and sparse-view CT reconstruction~\cite{sun2021coil}. In this work, we use INRs to approximate the signed distance functions of one or multiple surfaces. This INR is a fully connected multilayer perceptron (MLP), $f(\bm{x};\theta)$, that takes $\bm{x} \in \Omega := \left[-1, 1\right]^3 \subset \mathbb{R}^3$ as input, and outputs $SDF_\mathcal{M}(\bm{x})$, as shown in Figure \ref{fig:INR}\textit{(left)}. During optimization of the INR, desired properties of the SDF can be directly imposed on the network via the loss function. These properties include conditions on gradients, as they can be computed for the network through backpropagation \cite{sitzmann2019siren}. An INR thus represents a function or signal, and not a mapping between two functions, as is often the case with convolutional neural networks. As an INR is trained on continuous coordinates, they allow for representing a function at any resolution.

\subsection{Optimising an INR}
We use an INR to represent the SDF of a 3D vascular model, under the condition that the model should be easily obtainable from a 3D medical image volume. We start from a small set of points on the vessel surface: $\mathcal{X} = \{\bm{x}_i\}_{i=1,...,N} \subset \Omega$. We aim to simultaneously reconstruct an SDF with $\mathcal{X}$ on the zero level set and embed it in an INR. This amounts to solving the Eikonal equation with pointwise boundary conditions; an ill-posed problem. Gropp et al. \cite{gropp2020implicit} tackle this ill-posedness by minimising the following loss function:

\begin{align}\label{eq:loss_gropp}
    \ell(\theta) = \frac{1}{N}\sum\limits_{1 \leq i \leq N} \left(|f(\bm{x}_i;\theta)|\right) + \lambda \mathbb{E}_{x}\left(||\nabla_{\bm{x}} f(\bm{x};\theta)|| - 1 \right)^2.
\end{align}

The first term enforces that all points from $\mathcal{X}$ lie on the zero level set of the SDF, and thus on the surface $\mathcal{M}$. The second term is a regularising Eikonal term, encouraging that the INR $f(\bm{x};\theta)$ satisfies the Eikonal equation, just like the SDF it represents. The key observation of Gropp et al. is that minimising Eq. \eqref{eq:loss_gropp} yields an SDF with a smooth and plausible zero isosurface. This approach allows for flexibility in the consistency of $\mathcal{X}$, in contrast to similar methods \cite{mistelbauer2021implicit, schumann2007model}, that require local normals or points to lie within the same plane. In Section \ref{sec:robustness}, we investigate the quality and robustness of the reconstruction under variation of the point cloud size and locations.

\subsection{Fitting multiple functions in an INR}
The walls of blood vessels consist of layers, and in patients with, e.g., atherosclerosis or dissections it is important to model these layers separately. For example, in patients with abdominal aortic aneurysms (AAAs) the lumen and the surrounding solidified thrombus structure should be modeled separately~\cite{zhu2020intraluminal}. While INRs have mostly been used to represent single functions, they can be extended to fit multiple functions on the same domain. Figure \ref{fig:INR}\textit{(right)} visualizes how by adding additional output nodes to the neural network, multiple SDFs can be fitted simultaneously, in this case for the lumen, the inner wall, and outer wall of the AAA. This means that mutual properties of separate SDFs, such as for nested shapes, could be learned by a single neural network. This INR is optimised by minimising Eq. \eqref{eq:loss_gropp} on each of its output channels. We will compare this approach to multiple INRs representing structures separately in Section \ref{sec:nested}. % can be trained to simultaneously reconstruct the surfaces of each shape from their point cloud representation 

\subsection{Constructive solid geometry}
Signed distance functions make it very easy to determine the union, difference, or intersection of multiple shapes, a cumbersome and challenging task in explicit polygon mesh representations~\cite{jiang2016efficient, wu2010curvature}. Let $SDF_{\mathcal{M}_1}(\bm{x})$ and $SDF_{\mathcal{M}_2}(\bm{x})$ be SDFs for shapes $\mathcal{M}_1$ and $\mathcal{M}_2$ in $\Omega$, respectively. Then their union is determined as:
\begin{align}\label{eq:min}
    \mathcal{M}_1 \cup \mathcal{M}_2 = \min (SDF_{\mathcal{M}_1}(\bm{x}), SDF_{\mathcal{M}_2}(\bm{x}))
\end{align}
To allow for smoother blending between interfaces of shapes, a smoothed min function can be considered~\cite{li2007smooth}:
\begin{align}
    \begin{split} \label{eq:smoothblend}
    &\left( \mathcal{M}_1 \cup \mathcal{M}_2 \right)_{\text{smooth}} = \min(SDF_{\mathcal{M}_1}(\bm{x}), SDF_{\mathcal{M}_2}(\bm{x})) - \gamma_{\text{smooth}},\\
    &\gamma_{\text{smooth}} = \frac{1}{4k} \cdot \max(k - |SDF_{\mathcal{M}_1}(\bm{x}) - SDF_{\mathcal{M}_2}(\bm{x})|, 0)^2,
    \end{split}
\end{align}
where $k$ represents a smoothing parameter, i.e. a set region around the interfaces of the two surfaces. We will use this strategy to join structures represented by INRs in  Section \ref{section:cg}. %To compute the blended SDF of two shapes at location $\bm{x}$, we transform $\bm{x}$ to their respective domains $\Omega_{1,2}$ that each INR was trained on, forward them through the networks and apply equation \eqref{eq:smoothblend}.

\subsection{Data}\label{sec:data}
For all experiments conducted, we used 3D vascular models from two publicly available datasets. First, a dataset containing 19 3D geometries of abdominal aortic aneurysms \cite{wittek2020image}. Each of these geometries consists of three nested structures: lumen, inner- and outer wall. Second, the vascular model repository \cite{wilson2013vascular}, containing 119 mesh models of varying regions of the vascular system, designed for use in computational fluid dynamics.

For our experiments on robustness and nested shapes, we selected three different cases from the AAA dataset, with varying shape complexity. Case 1 contains a bifurcation and two major dilations of the aorta; the lumen mesh contains 11,490 vertices and 67,944 edges. Case 2 is a single vessel, with constant dilation; the mesh consists of 9,753 vertices and 57,387 edges. Similarly, Case 3 is a single vessel with no bifurcations, however there is some additional curvature in the center. The lumen mesh of Case 3 contains 9,816 vertices and 58,113 edges.

For the constructive geometry experiment, we randomly selected an aortofemoral tree from the vascular model repository. This geometry consisted of nine separate vessels, each of them given as a triangular mesh, as shown in Figure \ref{fig:combined_shapes}. The number of vertices and edges for these vessels varied between 8,608 and 70,284, and 51,636 and 421,692, respectively.

Since the INR operates on the domain $\Omega = [-1, 1]^3$, the data first requires some preprocessing to match this domain. For the nested AAA geometries, we rescaled the union of the vertices of the lumen, inner and outer wall to $\Omega$. For each of the three nested shapes, we randomly sampled a point cloud on their respective surfaces, that can be used for surface reconstruction using an INR. The original meshes are used to assess the quality of the surface reconstructions, using the average symmetric surface distance and Dice similarity coefficient.

For the meshes from the vascular model repository, we individually rescaled the mesh of each vessel to $\Omega$. Note that the physical domain each INR operates on hence differs between vessels. For each rescaled mesh, we randomly sampled a point cloud of a size proportional to the surface of the mesh, that is used to train the INR.

% \begin{figure}[t!]
%     \centering
%     \includegraphics[height=4cm]{Images/boxplots_DSC_ASD_lumen.png}
%     \caption{The Dice similarity coefficient (DSC) and average surface distance (ASD) of reconstructed vascular surface for varying number of reference points. Each SDF is reconstructed using 20 different point clouds of each size.}
%     \label{fig:boxplot_npoints}
% \end{figure}

\section{Experiments and Results}
We conducted three experiments: (1) assessing the robustness of the INR against variations in the point cloud, (2) testing the effectiveness of representing multiple shapes with a single INR and (3) demonstrating how our method can be used to reconstruct a vascular system from separate vessels and blend them naturally. For all experiments we used MLPs consisting of an input layer with three nodes, six fully connected hidden layers with 256 nodes and ReLU activations, and a final prediction layer consisting of one or three nodes, embedding a single shape or three nested shapes, respectively. Similar to \cite{gropp2020implicit}, we used a skip connection, connecting the input to the third hidden layer. We base our implementation on PyTorch code\footnote{\url{https://github.com/amosgropp/IGR}} provided with~\cite{gropp2020implicit}. 

\begin{figure}[t!]
    \centering
    \includegraphics[width=\textwidth, trim=0 0 6cm 0, clip]{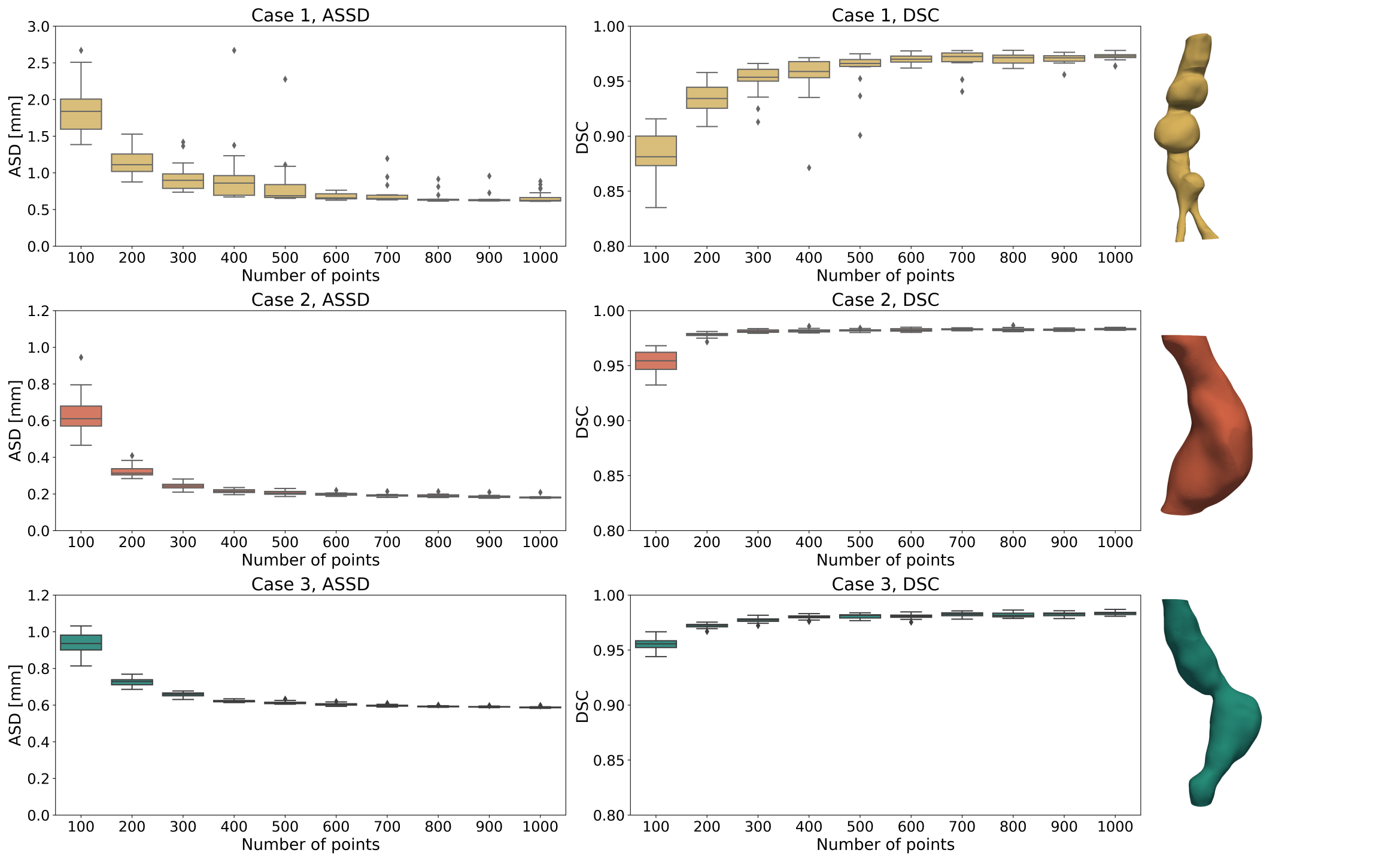}
    \caption{The Dice similarity coefficient (DSC) and average symmetric surface distance (ASSD) of reconstructed vascular surface for varying number of reference points. Each SDF is reconstructed using 20 different point clouds of each size.}
    \label{fig:boxplot_npoints}
\end{figure}

\subsection{Robustness}\label{sec:robustness}
We demonstrate how INRs can be integrated in the existing pipeline for vascular analysis. Instead of acquiring a tubular mesh by manual contour annotations, we fit an INR representation of the SDF to a point cloud. We assess the robustness against variations in this point cloud. For this experiment, we used three cases of the AAA dataset \cite{wittek2020image} as described in Section \ref{sec:data}. We sampled 20 different point clouds of predetermined sizes from the lumen mesh of each case. To train the INR reconstructing the geometries, we used an Adam optimizer with a learning rate of 0.0001, and set $\lambda$ in the loss function (Eq. \eqref{eq:loss_gropp}) to 0.1, as suggested in \cite{gropp2020implicit}. We trained the network for 25,000 epochs on an NVIDIA Quadro RTX 6000 GPU, taking 10-15 minutes per network.

Figure \ref{fig:boxplot_npoints} shows the quality of the surface reconstruction for each vascular structure, in terms of the Dice similarity coefficient (DSC) and average symmetric surface distance (ASSD). The DSC was computed using a binarized grid of the original mesh as a reference, the ASSD was computed based on the vertices of the original mesh. The median DSC for Case 2 and 3 is above 0.95 in all cases, even when only 100 points are used. For Case 1, which includes a bifurcation at the bottom, the median DSC at 100 points is slightly below 0.9. The median surface distances for Cases 2 and 3 are all sub-millimeter. The quality of the reconstruction of Case 1 lags compared to the other two, less complex, cases in terms of ASSD. For all three cases, we observe more variation in both ASSD and DSC when the reconstruction is based on fewer points. Medians of both metrics stabilize after 400 points for Case 2 and 3, and after 600 points for Case 1. Moreover, the standard deviations are small here, implying that the locations of the points will have a small impact on the final reconstructed shape.

\subsection{Reconstructing nested shapes}\label{sec:nested}
\begin{figure}[t!]
    \centering
    \includegraphics[width=\textwidth]{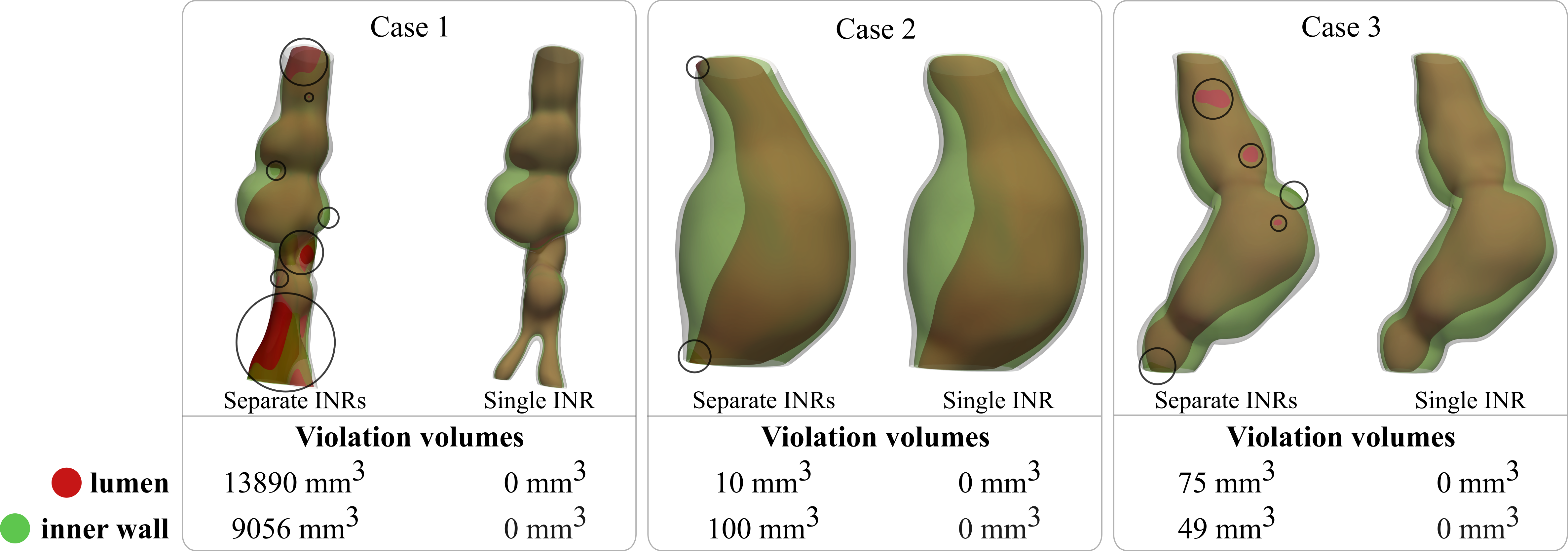}
    \caption{Three nested AAA structures represented by a single INR, or separate INRs each representing a single shape. Using a single INR preserves the nested property, violations of this anatomical prior are marked with circles.}
    \label{fig:nested_shapes}
\end{figure}

In our second experiment, we use the same AAA geometries as in Section \ref{sec:robustness}. However, we now use all three meshes of the vascular model: lumen, inner wall and outer wall of the aorta. Here, there is an anatomical prior that we want our model to capture. Namely, the lumen should be entirely nested inside the inner wall of the aorta, which is itself nested inside the outer wall. This implies: 
\begin{align}\label{SDF_property}
SDF_{\text{outer wall}}(\bm{x}) \leq SDF_{\text{inner wall}}(\bm{x}) \leq SDF_{\text{lumen}}(\bm{x}),\forall \bm{x} \in \Omega.
\end{align}
We used three AAA models from~\cite{wittek2020image}. For each, we sampled a point cloud of 200 points for each of the three surfaces; a proper trade-off between point cloud sparsity and reconstruction quality, per the results presented in Section \ref{sec:robustness}. We fit a joint INR to these point clouds, as well as three separate INRs, one for each of the three surfaces. Figure \ref{fig:nested_shapes} shows the resulting surfaces for both approaches. In all three cases, we see unwanted extrusions of the inner structures when separate INRs are used. Case 1 displays this effect most clearly; the surfaces of the lumen and inner wall (red and green structures respectively) fail to capture the bifurcation and are inconsistent with the outer wall. On the other hand, representation of the shapes obtained using a single INR consistently were free of such extrusions. Moreover, the thickness of the vessel wall remains constant for this approach, in agreement with ground-truth data. 

We quantified the unwanted extrusions in terms of violation volumes of the lumen and inner wall, also shown in Figure \ref{fig:nested_shapes}. These represent the volumes of the regions where each nested structure violates Eq. \eqref{SDF_property}. The violation volumes are defined as the volumes of the following sets:
\begin{align*}
    VV_{\text{lumen}} &= \{\bm{x} \in \Omega: SDF_{\text{lumen}}(\bm{x}) < 0 \land ( SDF_{\text{inner}}(\bm{x}) > 0 \lor SDF_{\text{outer}}(\bm{x}) > 0 ) \} \\
    VV_{\text{inner}} &= \{\bm{x} \in \Omega: SDF_{\text{inner}}(\bm{x}) < 0 \land SDF_{\text{outer}}(\bm{x}) > 0\}.
\end{align*}
The values reported in Figure \ref{fig:nested_shapes} are the volumes scaled back to physical sizes. We observe that using separate INRs for the reconstruction of Case 1, the most complex geometry, results in extremely high violation volumes compared to the other two cases. As we observed in our qualitative assessment, unwanted extrusions do consistently not appear when separate INRs are used.

%We repeated this experiment for point clouds of size 500, 750 and 1000 and observed that the inner structures represented by separate INRs showed fewer extrusions.

\subsection{Constructive geometry}\label{section:cg}
\begin{figure}[t!]
    \centering
\includegraphics[width=\textwidth]{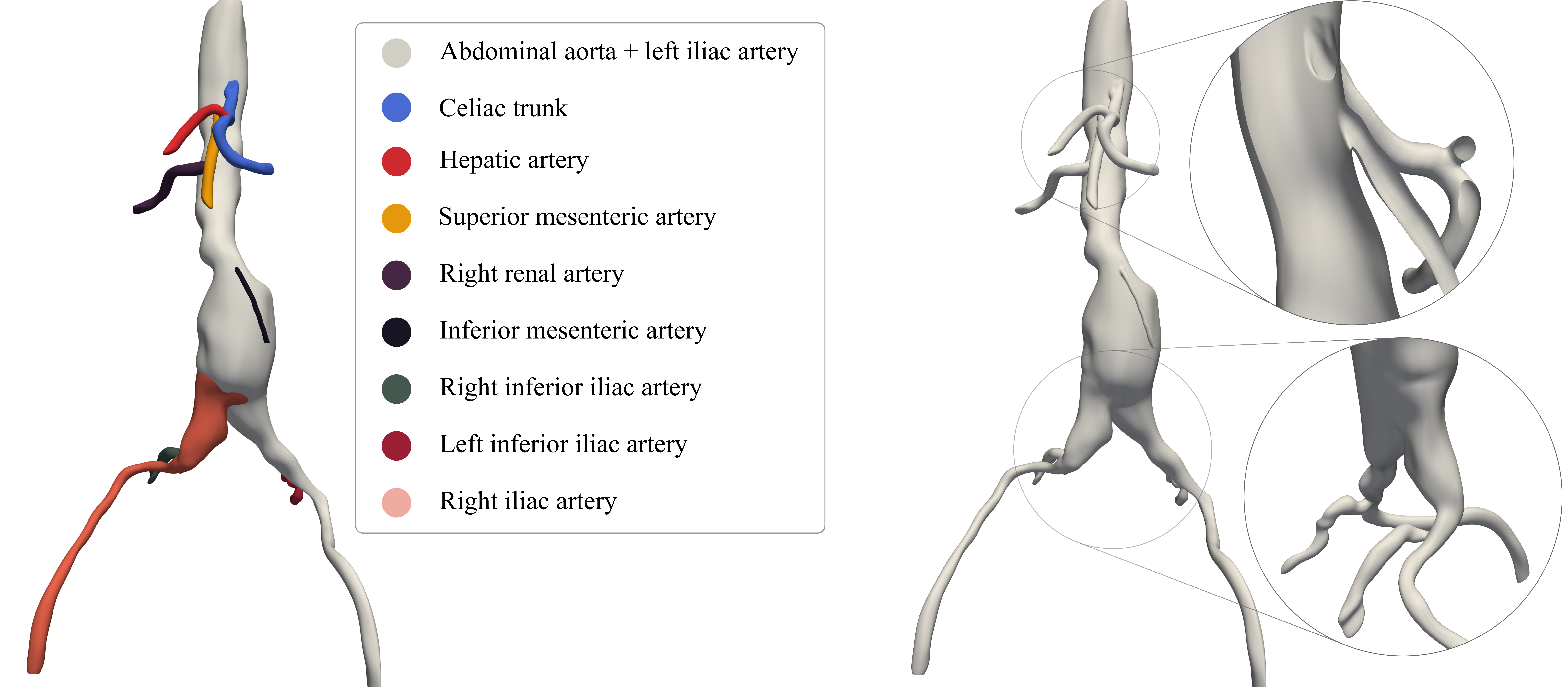}
\caption{Subdomains of separate INRs, indicated by different colours (left) and smooth blending of these INRs, with an inside-view at the top (right).}
    \label{fig:combined_shapes}
\end{figure}

In our third experiment, we demonstrate how INRs enable easy and smooth surface blending. This is critical for personalised vascular modeling, where many arteries of different calibers need to be blended into one single model for, e.g., CFD. Here, we use an aortofemoral tree from~\cite{wilson2013vascular}, that we preprocessed as described in Section \ref{sec:data}. This sample consists of nine separate mesh structures: aorta, superior and inferior mesenteric arteries (SMA/IMA), renal arteries, hepatic artery, and common and internal iliac arteries, shown in Figure \ref{fig:combined_shapes}\textit{(left)}. We fit an INR to a point cloud of each of the vessels. %, after normalizing it to the $\left[-0.9, 0.9 \right]^3$ domain to allow for smooth blending. 
We optimize these networks using an Adam optimizer for 50,000 epochs. % We also re-weighted the first term in the loss function from 1 to 5.

To blend the vessels, we generate a Cartesian grid encompassing the entire vascular tree. Hence, this domain contains the respective subdomains that all nine INRs were trained on. To evaluate the INRs for each structure on this grid, we map each grid point to the corresponding coordinate in each INR domain and query the SDF value by forwarding it through the network. Next, we rescale the SDF value from the INR domain to the real size. This results in a grid containing patches of SDFs, that we blend together using Eq. \eqref{eq:smoothblend}, with smoothing parameter $k=0.1$. The result in Figure \ref{fig:combined_shapes}\textit{(right)} shows a smooth transition between the different vascular structures, without any intrusions on the inside, making this structure suitable for CFD analysis\cite{kretschmer2013interactive}. %Planar caps at the end of the vessels, used to impose boundary conditions, can be obtained by centerline extraction using the SDF.

% \begin{itemize}
%     \item Evaluated the average surface distance as well as the Dice Similarity Coefficient for the reconstructed shapes, compared to the original mesh annotations
%     \item Decent reconstructions of the shapes, even with as little as 100 points to reconstruct the surface
%     \item The variation in reconstruction quality is big for smaller pointclouds, but decreases to nearly 0 if more points are included.
%     \item The lumen is a lot 'lumpier' than the thrombus internal and external
%     \item Additional loss term has little effect on the reconstructed surface in terms of the Dice and ASD, but it does have effect on the separate SDF's if their shapes differ a lot. (A lot of SDF violation was observed outside the shapes between the lumen and the internal wall of the thrombus). 
% \end{itemize}

\section{Discussion and Conclusion}
In this work, we have shown the value of implicit neural representations for 3D vascular modeling. We have demonstrated that INRs can be used to fit surfaces in abdominal aortic aneurysms using a small number of annotated points on the surface. Moreover, results show that fitting multiple nested surfaces in one INR is less likely to lead to violation of anatomical priors. Finally, we have used simple arithmetics to blend multiple surfaces represented in INRs into one complex vascular model.

Following the work of \cite{gropp2020implicit}, we have shown that we can reconstruct continuous surfaces based on (small) point clouds, where their implicit prior leads to surfaces that are smooth and interpolate well in space. This contrasts with, e.g., slice interpolation as is often used to obtain dense voxel masks. INRs are \textit{continuous}, allowing us to represent the vessel surface at any arbitrary resolution. This could facilitate seamless blending of separate arteries for, e.g., CFD analysis \cite{kretschmer2013interactive}. 

We have demonstrated how a single INR can simultaneously parametrise multiple nested surfaces in an abdominal aortic aneurysm. We showed that using a single INR to represent nested shapes benefits the consistency of the surfaces. The SDF values of nested shapes are similar, as they differ by a local offset. Using a single network to reconstruct them enables the network to learn these offsets by sharing weights and information from the other shapes during training. This nested property is, however, not a topological guarantee, but can be enforced by additional regularisation in the loss function.

Explicit representations such as polygonal meshes ~\cite{antiga2008image, lan2018re, arthurs2021crimson} and voxel masks are ubiquitous in vascular modeling. A direct comparison in this work between explicit representations and INRs is challenging, as we used triangular meshes as ground-truth geometries. However, we argue that instead of replacing explicit representations with INRs, they can cooperate. We demonstrated how INRs can be obtained from an explicit representation, and that smoothly combining them is straightforward. Afterwards, a mesh representation of the combined shape can be obtained at any desired resolution, obviating the need of cumbersome mesh operations~\cite{jiang2016efficient}.

One limitation of INRs is their lack of generalisation; a new INR needs to be optimised for each shape. This may be overcome by enriching the network input with a latent code, in which case INRs could represent a whole distribution of shapes~\cite{park2019deepsdf}. Alternatively, INRs could be embedded in existing deep learning methods for (cardiovascular) image segmentation, e.g., by including local image features~\cite{chen2021learning} or training hypernetworks~\cite{sitzmann2019metasdf}. Also, our work could lead to interactive annotation tools. The combination of recently proposed rapid INR fitting approaches~\cite{mueller2022instant} and uncertainty quantification~\cite{gal2016dropout} could guide the annotator to areas that need additional annotations.

Besides their use in manual annotation procedures, INRs can be used in addition to automatic segmentation of vascular structures. For example, in \cite{Alblas/et/al/2021} contour points of the lumen and outer wall of atherosclerotic carotid arteries are automatically acquired from MR images. These points can be used for optimizing the INR and acquiring a continuous surface representation of the nested lumen and outer wall of the carotid arteries.

% These methods have shown to be successful in processing image information, hence their image encodings are very expressive. These image encodings may be used to condition an INR, integrating continuous shape representations and image information. To make the model more robust against widely varying anatomy and pathology among patients, local image features may be used \cite{chen2021learning}.

In conclusion, INRs are a versatile surface representation that are easily acquired and integrated in existing frameworks for vascular modeling.

%%% JMW 22-01
% Ook noemen
% \begin{itemize}
%     \item Dat je de markers die je zet in een artbitrair vlak gezet kunnen worden (dus niet per se in axiaal, sagittaal of coronal)
%     \item Dat je de markers maximaal benut omdat ze allemaal onafhankelijk iets bij dragen aan de 3D shape
%     \item Waarom je voor een implicit representation gaat. Waarom niet voor een voxel masker of voor een triangular mesh? Je kunt zeggen dat je in alle gevallen interpolatie moet doen. De surface van je structuur is een manifold in 3D en je hebt alleen maar gesamplede waarden op die manifold (je markers/point cloud). De meest volledige manier om de manifold te reconstrueren is door de volledige impliciete omschrijving ervan te vinden. Ook hier verwijzen naar die radial basis functions en spherical harmonics (\url{https://www.hindawi.com/journals/ijbi/2011/658930/}).
%     \item Ook goed om te benoemen dat (met name voor vaten) er sowieso al veel interpolatie wordt gedaan op basis van splines en NURBS. Dat is natuurlijk ook weer gerelateerd aan radial basis functions, spherical harmonics, waar je bepaalde basis functions gebruikt. Als je geen basis functions gebruikt maar een MLP heb je veel meer vrijheidsgraden. Misschien is het goed om expliciet het onderscheid te maken tussen \textbf{hand-crafted} parametrizations zoals NURBS, splines, RBF, spherical harmonics, en \textbf{learned} parametrization van je vorm in een MLP.
% \end{itemize}

% ---- Bibliography ----
%
% BibTeX users should specify bibliography style 'splncs04'.
% References will then be sorted and formatted in the correct style.
%
\bibliographystyle{splncs04}
\bibliography{Going_offgrid}

\begin{thebibliography}{10}
\providecommand{\url}[1]{\texttt{#1}}
\providecommand{\urlprefix}{URL }
\providecommand{\doi}[1]{https://doi.org/#1}

\bibitem{Alblas/et/al/2021}
Alblas, D., Brune, C., Wolterink, J.M.: {Deep Learning-Based Cartotid Artery
  Vessel Wall Segmentation in Black-Blood MRI Using Anatomical Priors}. In:
  SPIE Medical Imaging (2022)

\bibitem{antiga2008image}
Antiga, L., Piccinelli, M., Botti, L., Ene-Iordache, B., Remuzzi, A., Steinman,
  D.A.: An image-based modeling framework for patient-specific computational
  hemodynamics. {MBEC}  \textbf{46}(11),  1097--1112 (2008)

\bibitem{arthurs2021crimson}
Arthurs, C.J., Khlebnikov, R., Melville, A., Mar{\v{c}}an, M., Gomez, A.,
  Dillon-Murphy, D., Cuomo, F., Silva~Vieira, M., Schollenberger, J., Lynch,
  S.R., et~al.: {CRIMSON: An open-source software framework for cardiovascular
  integrated modelling and simulation}. {PLoS Comp. Biol.}  \textbf{17}(5),
  e1008881 (2021)

\bibitem{van2002level}
van Bemmel, C.M., Spreeuwers, L.J., Viergever, M.A., Niessen, W.J.: Level-set
  based carotid artery segmentation for stenosis grading. In: International
  Conference on MICCAI. pp. 36--43. Springer (2002)

\bibitem{carr1997surface}
Carr, J.C., Fright, W.R., Beatson, R.K.: Surface interpolation with radial
  basis functions for medical imaging. {IEEE Trans Med Imaging}
  \textbf{16}(1),  96--107 (1997)

\bibitem{chen2020deep}
Chen, C., Qin, C., Qiu, H., Tarroni, G., Duan, J., Bai, W., Rueckert, D.: Deep
  learning for cardiac image segmentation: a review. Front. cardiovasc. med.
  \textbf{7}, ~25 (2020)

\bibitem{chen2021learning}
Chen, Y., Liu, S., Wang, X.: Learning continuous image representation with
  local implicit image function. In: CVPR. pp. 8628--8638. IEEE/CVF (2021)

\bibitem{gal2016dropout}
Gal, Y., Ghahramani, Z.: {Dropout as a Bayesian approximation: Representing
  model uncertainty in deep learning}. In: {ICML}. pp. 1050--1059. PMLR (2016)

\bibitem{gansca2002self}
Gansca, I., Bronsvoort, W.F., Coman, G., Tambulea, L.: Self-intersection
  avoidance and integral properties of generalized cylinders. Comput. Aided
  Geom. Des.  \textbf{19}(9),  695--707 (2002)

\bibitem{gropp2020implicit}
Gropp, A., Yariv, L., Haim, N., Atzmon, M., Lipman, Y.: Implicit geometric
  regularization for learning shapes. In: ICML. pp. 3789--3799 (2020)

\bibitem{hong2020high}
Hong, Q., Li, Q., Wang, B., Tian, J., Xu, F., Liu, K., Cheng, X.: High-quality
  vascular modeling and modification with implicit extrusion surfaces for blood
  flow computations. Comput Methods Programs in Biomed  \textbf{196},  105598
  (2020)

\bibitem{jiang2016efficient}
Jiang, X., Peng, Q., Cheng, X., Dai, N., Cheng, C., Li, D.: Efficient booleans
  algorithms for triangulated meshes of geometric modeling. Comput. Aided Des.
  Appl.  \textbf{13}(4),  419--430 (2016)

\bibitem{kretschmer2013interactive}
Kretschmer, J., Godenschwager, C., Preim, B., Stamminger, M.: Interactive
  patient-specific vascular modeling with sweep surfaces. IEEE Trans Vis Comput
  Graph  \textbf{19}(12),  2828--2837 (2013)

\bibitem{lan2018re}
Lan, H., Updegrove, A., Wilson, N.M., Maher, G.D., Shadden, S.C., Marsden,
  A.L.: {A re-engineered software interface and workflow for the open-source
  SimVascular cardiovascular modeling package}. {J. Biomech. Eng.}
  \textbf{140}(2) (2018)

\bibitem{li2007smooth}
Li, Q.: Smooth piecewise polynomial blending operations for implicit shapes.
  In: Comput Graph Forum. vol.~26, pp. 157--171. Wiley Online Library (2007)

\bibitem{litjens2019state}
Litjens, G., Ciompi, F., Wolterink, J.M., de~Vos, B.D., Leiner, T., Teuwen, J.,
  I{\v{s}}gum, I.: State-of-the-art deep learning in cardiovascular image
  analysis. JACC: Cardiovasc. imaging  \textbf{12}(8 Part 1),  1549--1565
  (2019)

\bibitem{lorigo2001curves}
Lorigo, L.M., Faugeras, O.D., Grimson, W.E.L., Keriven, R., Kikinis, R.,
  Nabavi, A., Westin, C.F.: {Curves: Curve evolution for vessel segmentation}.
  {Med. Image Anal.}  \textbf{5}(3),  195--206 (2001)

\bibitem{martel2021acorn}
Martel, J.N., Lindell, D.B., Lin, C.Z., Chan, E.R., Monteiro, M., Wetzstein,
  G.: Acorn: adaptive coordinate networks for neural scene representation. ACM
  Trans. Graph.  \textbf{40}(4),  1--13 (2021)

\bibitem{mildenhall2020nerf}
Mildenhall, B., Srinivasan, P.P., Tancik, M., Barron, J.T., Ramamoorthi, R.,
  Ng, R.: Nerf: Representing scenes as neural radiance fields for view
  synthesis. In: ECCV. pp. 405--421. Springer (2020)

\bibitem{mistelbauer2021implicit}
Mistelbauer, G., R{\"o}ssl, C., B{\"a}umler, K., Preim, B., Fleischmann, D.:
  Implicit modeling of patient-specific aortic dissections with elliptic
  fourier descriptors. In: Comput Graph Forum. vol.~40, pp. 423--434. Wiley
  Online Library (2021)

\bibitem{mueller2022instant}
M\"uller, T., Evans, A., Schied, C., Keller, A.: {Instant Neural Graphics
  Primitives with a Multiresolution Hash Encoding}. arXiv:2201.05989  (2022)

\bibitem{park2019deepsdf}
Park, J.J., Florence, P., Straub, J., Newcombe, R., Lovegrove, S.: {DeepSDF:
  Learning Continuous Signed Distance Functions for Shape Representation}. In:
  CVPR. pp. 165--174. IEEE/CVF (2019)

\bibitem{schumann2007model}
Schumann, C., Oeltze, S., Bade, R., Preim, B., Peitgen, H.O.: Model-free
  surface visualization of vascular trees. In: EuroVis. pp. 283--290 (2007)

\bibitem{shani1984splines}
Shani, U., Ballard, D.H.: Splines as embeddings for generalized cylinders.
  Comput. Vis., Graph, and Image Process.  \textbf{27}(2),  129--156 (1984)

\bibitem{sitzmann2019metasdf}
Sitzmann, V., Chan, E.R., Tucker, R., Snavely, N., Wetzstein, G.: {MetaSDF:
  Meta-Learning Signed Distance Functions}. In: {NeurIPS} (2020)

\bibitem{sitzmann2019siren}
Sitzmann, V., Martel, J.N., Bergman, A.W., Lindell, D.B., Wetzstein, G.:
  {Implicit Neural Representations with Periodic Activation Functions}. In:
  NeurIPS (2020)

\bibitem{sobocinski2013benefits}
Sobocinski, J., Chenorhokian, H., Maurel, B., Midulla, M., Hertault, A.,
  Le~Roux, M., Azzaoui, R., Haulon, S.: {The benefits of EVAR planning using a
  3D workstation}. Eur J Vasc Endovasc Surg  \textbf{46}(4),  418--423 (2013)

\bibitem{sun2021coil}
Sun, Y., Liu, J., Xie, M., Wohlberg, B., Kamilov, U.S.: Coil: Coordinate-based
  internal learning for tomographic imaging. IEEE Trans Comput Imaging
  \textbf{7},  1400--1412 (2021)

\bibitem{swart2019shared}
Swart, M., McCarthy, R.: Shared decision making for elective abdominal aortic
  aneurysm surgery. Clinical Medicine  \textbf{19}(6), ~473 (2019)

\bibitem{tran2021patient}
Tran, K., Yang, W., Marsden, A., Lee, J.T.: Patient-specific computational flow
  modelling for assessing hemodynamic changes following fenestrated
  endovascular aneurysm repair. JVS: Vascular Science  \textbf{2},  53--69
  (2021)

\bibitem{wilson2013vascular}
Wilson, N.M., Ortiz, A.K., Johnson, A.B.: The vascular model repository: a
  public resource of medical imaging data and blood flow simulation results. J.
  Med. Devices  \textbf{7}(4) (2013)

\bibitem{wittek2020image}
Wittek, A., Mufty, H., Catlin, A., Rogers, C., Saunders, B., Sciarrone, R.,
  Fourneau, I., Meuris, B., Tavner, A., Joldes, G.R., et~al.: Image, geometry
  and finite element mesh datasets for analysis of relationship between
  abdominal aortic aneurysm symptoms and stress in walls of abdominal aortic
  aneurysm. {Data Br.}  \textbf{30},  105451 (2020)

\bibitem{wolterink2021implicit}
Wolterink, J.M., Zwienenberg, J.C., Brune, C.: {Implicit Neural Representations
  for Deformable Image Registration}. In: MIDL. PMLR (2022)

\bibitem{wu2010curvature}
Wu, J., Ma, R., Ma, X., Jia, F., Hu, Q.: Curvature-dependent surface
  visualization of vascular structures. {Computerized Medical Imaging and
  Graphics}  \textbf{34}(8),  651--658 (2010)

\bibitem{zhu2020intraluminal}
Zhu, C., Leach, J.R., Wang, Y., Gasper, W., Saloner, D., Hope, M.D.:
  Intraluminal thrombus predicts rapid growth of abdominal aortic aneurysms.
  Radiology  \textbf{294}(3),  707--713 (2020)

\end{thebibliography}
\end{document}